\documentclass[prb,twocolumn,showpacs,amsmath,amssymb]{revtex4}

\usepackage{graphicx}% Include figure files
\usepackage{dcolumn}% Align table columns on decimal point
\usepackage{bm}% bold math
\usepackage[dvips]{color}
\usepackage{color}

\newcommand{\vect}[1]{\mathbf{#1}}
\newcommand{\figwidth}{0.9\columnwidth}

\begin{document}
\preprint{APS/123-QED}

\title{ESR of the quasi-two-dimensional antiferromagnet CuCrO$_2$ with a triangular
    lattice}

\author{A.~M. Vasiliev}
\author{L.~A. Prozorova}
\author{L.~E. Svistov}
\email{svistov@kapitza.ras.ru}
\affiliation{P. L. Kapitza Institute for
Physical Problems RAS, 119334 Moscow, Russia}

\author{V. Tsurkan}
\affiliation{Institute of Applied Physics, Academy of Sciences of
Moldova, MD--2028 Chisinau, Republic of Moldova }

\affiliation{Center for Electronic Correlations and Magnetism,
Experimentalphysik V, Universit\"{a}t Augsburg, D--86135 Augsburg,
Germany}

\author{V. Dziom}
\author{A. Shuvaev}
\author{Anna Pimenov}
\author{A. Pimenov}
%\email{pimenov@ifp.tuwien.ac.at}
\affiliation{Institute of Solid State Physics, Vienna University of Technology, A-1040 Vienna, Austria}

\date{\today}

\begin{abstract}

Using electron-spin-resonance (ESR) technique we investigate the
magnetic structure of CuCrO$_2$, quasi-two-dimensional
antiferromagnet with weakly distorted triangular lattice. Resonance
frequencies and the excitation conditions in CuCrO$_2$ at low
temperatures are well described in the frame of cycloidal spin
structure, defined by two susceptibilities parallel and
perpendicular to the spin plane ($\chi_{\perp}$ and
$\chi_{\parallel}$) and by a biaxial crystal-field anisotropy. In
agreement with the calculations, the character of the eigenmodes
changes drastically at the spin-flop transition. The splitting of
the observed modes can be well attributed to the resonances from
different domains. The domain structure in CuCrO$_2$ can be
controlled by annealing of the sample in magnetic field.

\end{abstract}

\maketitle

\section{Introduction}

Magnetic materials with frustrated exchange interactions are one of
the attractive issues in modern solid state physics. In these
materials unconventional magnetic orders appear as a subtle balance
of exchange energies and they are often governed by much weaker
interactions or fluctuations. Such frustrated systems are known in
nature for quasi-one-dimensional, quasi-two-dimensional and for
three-dimensional cases. For one(two)-dimensional magnets the
interactions within the chain (plane) are much larger then the
coupling of spins from different chains (plains).

Magnetic properties of an antiferromagnet on a regular triangular
planar lattice have been intensively studied
theoretically.~\cite{Kawamura_1985,Korshunov_1986,Anderson_1987,%
Plumer_1990,Chubukov_1991,White_2007}
The ground state in the Heisenberg and XY-models is a ``triangular''
planar spin structure with three magnetic sublattices arranged by
120$^\circ$ apart.
%The orientation of the
%spin plane is not fixed in the exchange approximation in the Heisenberg model.
%Quantum fluctuations in such systems reduce the moments of the magnetic ions
%essentially.
After neutron scattering experiments~\cite{Kadowaki_1990} CuCrO$_2$
was for a long time ascribed to magnets with regular triangular
structure and with 120$^\circ$ spin arrangement in the planes. A
disorder caused by frustration of the inter-plane exchange bonds was
also suggested. Recent neutron scattering
investigations~\cite{Poeinar_2009} in CuCrO$_2$ single crystals
detected a three dimensional magnetic order with incommensurate wave
vector that slightly differs from the wave vector of  a commensurate
120$^\circ$ structure. The magnetic ordering is accompanied by a
simultaneous crystallographic distortion~\cite{Kimura_2009} of the
regular triangular lattice and by the appearance of an electrical
polarization. The electric polarization in CuCrO$_2$ can be
influenced by comparatively weak magnetic field. Microscopic
mechanism of this complex magneto-elastic transition is not well
understood and is a subject of actual investigations.

Electron spin resonance (ESR) technique is a powerful tool to
investigate magnetically ordered states of matter. The high field
ESR in CuCrO$_2$ was investigated in
Ref.~[\onlinecite{Yamaguchi_2010}]. Although the flop of the
cycloidal plane was detected at $H_c = 5.3$~T, many important issues
such as the polarization analysis of the excitation conditions and
the separation of the contributions from different domains remain
open. In this work the detailed magnetic and domain dynamics of
quasi-two dimensional antiferromagnet with the triangular lattice
CuCrO$_2$ has been investigated using ESR technique in a broad
frequency range.

\section{Crystal structure and magnetism in C$\rm{u}$C$\rm{r}$O$_2$}

%\begin{figure}
%\includegraphics[width=\figwidth,angle=0,clip]{struct.eps}
%\caption{.} \label{fig:struct}
%\end{figure}

\begin{figure}
\includegraphics[width=\figwidth,angle=0,clip]{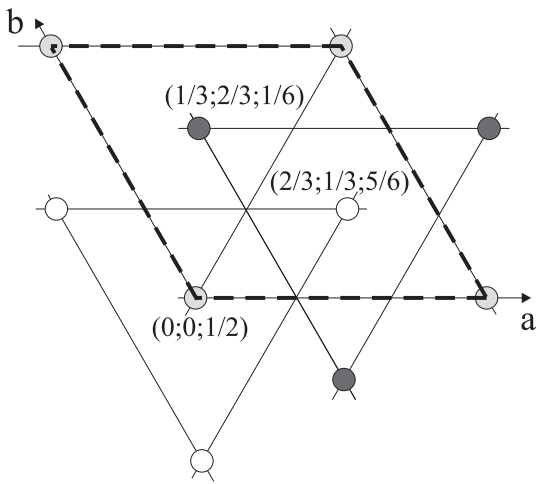}
\caption{Crystal structure of CuCrO$_2$ in projection on the
$ab$-plane. The positions of Cr$^{3+}$ ions are marked by circles. The
crystal cell is outlined with dashed line.} \label{fig0}
\end{figure}

CuCrO$_2$ crystallizes in a delafossite structure (space group
$R\bar{3}m$) with the following  hexagonal unit cell parameters at
room temperature: $a=2.98$~\AA{}, $c =17.11$~\AA{}. The unit cell of
CuCrO$_2$ contains three formula units. Chromium ions occupy the
positions (0; 0; 1/2), (1/3; 2/3; 1/6), (2/3; 1/3; 5/6) in the
crystal cell (see Fig.~\ref{fig0}) and Cu$^{+}$ ions occupy the positions
(0; 0; 0), (1/3; 2/3; 2/3), (2/3; 1/3; 1/3) (Ref.~[\onlinecite{Beznos_2009}]).
Magnetic Cr$^{3+}$ ions ($S=3/2$) form a triangular lattice in the
$ab$-planes. Adjacent planes are separated by nonmagnetic copper
ions along the $c$-axis. At temperatures above the N\'{e}el
temperature $(T > T_N\approx 24$ K) the triangular lattice is
regular. In the magnetically ordered state the triangular lattice is
distorted, such that one side of the triangle becomes slightly
smaller then two other sides: $\Delta a / a \simeq
10^{-4}$~(Ref.~[\onlinecite{KimuraJ_2009}]).

According to neutron diffraction experiments at low temperatures a
spiral incommensurate spin structure with a wave vector
$\vect{q}_{ic}= (0.329, 0.329, 0)$ is
established~\cite{Poeinar_2009,Poeinar_2010,Frontzek_2012}. The
cycloidal rotation of the magnetic moments $\vect{M}_i$ of Cr$^{3+}$
ions may be written as:
\begin{eqnarray}
\vect{M}_i=M_1\vect{e}_1\cos(\vect{q}_{ic}\cdot\vect{r}_i+\psi)+M_2\vect{e}_2\sin(\vect{q}_{ic}\cdot\vect{r}_i+\psi),
\label{eqn:magn}
\end{eqnarray}
where $\vect{e}_1$ and $\vect{e}_2$ are two perpendicular unit vectors
determining the spin plane orientation with the normal vector
$\vect{n}=\vect{e}_1 \times \vect{e}_2$, and $\psi$ is an arbitrary phase. The
values of the magnetic components ($M_1$, $M_2$) depend on the arrangement of
the spiral plane with respect to the crystal axes. For zero magnetic field
$\vect{e_1}$ is parallel to $[\bar{1}10]$ with $M_1 = 2.2(2)$ $\mu_B$ and
$\vect{e_2}$ is parallel to $[001]$ with $M_2 = 2.8(2)$ $\mu_B$,
respectively (Ref.~[\onlinecite{Frontzek_2012}]). The difference between $M_1$
and $M_2$ reflects the ellipticity of the helix.

According to the results of inelastic neutron scattering~\cite{Poeinar_2010}
the interaction is strongest in the trigonal plane between the nearest
Cr$^{3+}$ ions with the exchange constant $J_{ab}=2.3$ meV. The inter-plane
interaction is frustrated and at least one order of magnitude weaker than the
in-plane interaction. The nature of the incommensurability of magnetic
structure is not clear at present. The value of the incommensurate vector
$(0.329,0.329,0)$ is very close to the vector of the regular triangle structure
$(1/3,1/3,0)$ and can be explained by weak interactions. For example, the
incommensurability can be influenced by small difference of in-plane
interactions of distorted triangles: $J_{ab}/J'_{ab}=-2\cos(2\pi q_{ic})\approx
1.05$. Here $J'_{ab}$ is the exchange interaction along the nonequal edges of
the distorted triangle. Alternatively, the observed incommensurability can be
explained by joint influence of the inter-plane and the next-nearest
intra-plane interactions~\cite{Poeinar_2010}.

The plane of the spin cycloid in CuCrO$_2$ is perpendicular to one
side of the triangle~\cite{Soda_2009}. Accordingly, three equivalent
magnetic domains coexist in the ordered state. Such arrangement of
the spin plane is in agreement with a strong ``easy axis'' single
ion anisotropy along the $c$-axis as obtained from inelastic neutron
scattering experiments~\cite{Poeinar_2010}. Therefore, the magnetic
domains are characterized by both, the distortion of the triangular
plane parallel to [100], [010], [110] and the orientation of the
cycloidal spin plane with the normal vector
$\vect{n}_{1,2,3}\parallel [100], [010], [110]$, respectively. Such
domains will be referred in the text as [100], [010] and [110]
domains.

The orientation of the spin plane with respect to triangular plane
is defined by a weak in-plane anisotropy and can be influenced by
moderate external magnetic field. For $\vect{H}\parallel [1\bar{1}0]$
the spin reorientation transition was observed at
$5.3$~T.~\cite{Yamaguchi_2010}

\section{Experimental details}

Single crystals of CuCrO$_2$ were grown by a flux method in Pt
crucibles using Bi$_2$O$_3$ as a solvent and starting with
polycrystalline pellets. Powder X-ray diffraction measurements of
the single crystals did not show any
%detectable
impurity phases. The crystals had a platelet
shape (approximately $3\times3\times0.5$~mm$^3$) with the large
surface perpendicular to the hexagonal $c$-axis.

Initial polycrystalline CuCrO$_2$ samples were prepared by solid-state reaction
from a stoichiometric mixture of CuO and Cr$_2$O$_3$ \cite{Seki_2008}. The
mixture was pressed into pellets and sintered at 1000$^{\circ}$C for 40 hours
in air. This procedure was repeated after intermediate grinding for 40~h at the
same temperature.

The ESR experiments were performed with a transmission-type
spectrometer using various resonators in the frequency range
$14<\omega/2\pi<140$~GHz. The superconducting solenoid has provided
magnetic fields up to $8$~T.

The high frequency branch of the spectra was studied using the
quasi-optical technique~\cite{volkov_infrared_1985,%
pimenov_prb_2005}. In the case of magnetic excitations the
transmission through the sample can be obtained as a function of
either temperature, frequency, or external magnetic
field~\cite{ivannikov_prb_2002}. Due to well-defined polarization of
the electromagnetic radiation the orientation of the ac magnetic
field is also known in addition to the direction of the static
field. For the present study the following ranges of external
parameters have been utilized: $300 <\omega/2\pi< 450$~GHz,
$0<H<7$~T, and $2<T<30$~K. The spectra have been analyzed using
Fresnel optical expressions for the transmission of a plane-parallel
sample assuming Lorentzian form of the complex magnetic
permeability:
\begin{equation}\label{mu}
    \mu^{\ast}(H)=1+\frac{\Delta\mu H_0^2}{H_0^2-H^2-iH\delta} \, .
\end{equation}
Here $\Delta \mu$, $H_0$, and $\delta$ are magnetic contribution,
resonance field and resonance width, respectively.

\section{Results}

\begin{figure}
\includegraphics[width=.9\columnwidth,angle=0,clip]{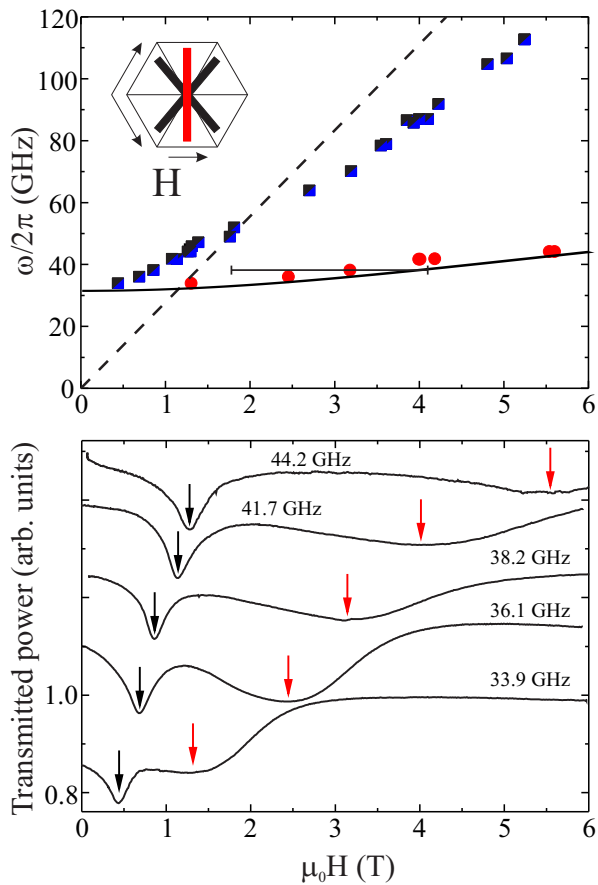}
\caption{(color online) Upper panel: Magnetic field dependence of
the ESR frequencies for CuCrO$_2$ single crystal. Magnetic field is
directed parallel to one side of the triangular structure:
$H\parallel[110]$. Red circles can be attributed to the ESR from the
domains with the distorted side aligned along the $[110]$ direction.
The black-blue squares corresponds to resonances from the domains
with the distortions along $[100]$ and $[010]$, respectively. The
geometry of the experiment is shown in the inset. Solid line shows
the calculated $\omega(H_{res})$ within the model discussed in the
text. The dotted line corresponds to a paramagnetic mode with
$g$-factor equal 2. Lower panel: Examples of ESR absorption spectra
at $T=4.2$~K. Black and red arrows mark the absorption modes
corresponding to the black and red symbols on the top panel.}
\label{fig1}
\end{figure}

\begin{figure}
\includegraphics[width=.9\columnwidth,angle=0,clip]{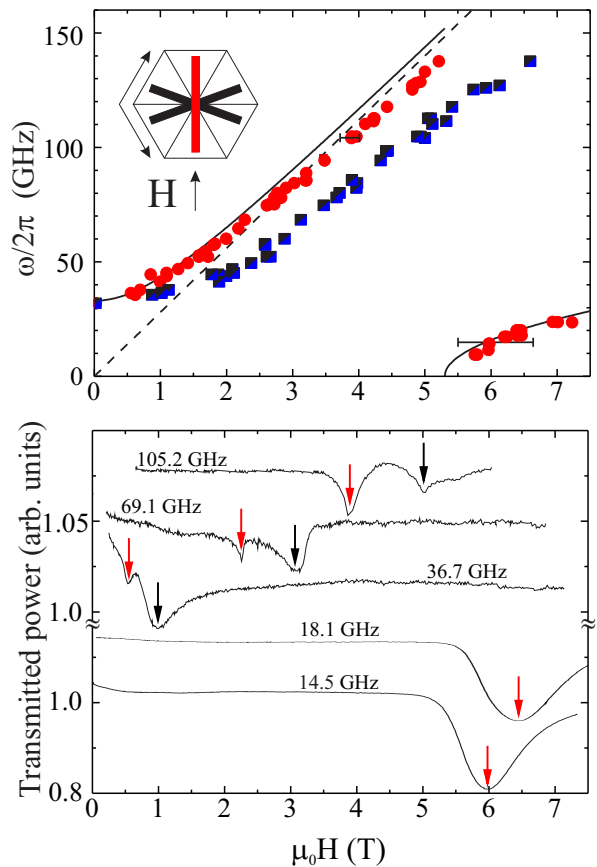}
\caption{(color online)  Upper panel: ESR frequencies in CuCrO$_2$
for magnetic field directed perpendicular to one side of the
triangular structure: $H\parallel[\bar{1}10]$. Red circles can be
attributed to the modes from the domains with the distorted side
aligned along $[110]$. The black-blue squares corresponds to
resonances from the domains with distortions along [100] and [010].
Orientations of the crystallographic axes, applied field and spin
planes of three domains are shown in the inset. Solid lines show the
calculated $\omega(H_{res})$ dependences within the model discussed
in the text. The dotted line corresponds to a paramagnetic mode with
$g=2$. Lower panel: Examples of ESR absorption lines at $T=4.2$~K.
Black and red arrows mark the absorption lines corresponding to the
black and red symbols on the top.} \label{fig2}
\end{figure}

Lower panels of Figs.~\ref{fig1},~\ref{fig2} show magnetic field dependencies
of the ESR signal in CuCrO$_2$ at different radiation frequencies. These curves
were obtained for the magnetic field directed parallel (Fig.~\ref{fig1}) and
perpendicular (Fig.~\ref{fig2}) to one side of the triangular lattice.
Corresponding frequencies of the resonance field ($H_{res}$) are shown in the
upper panels. Since the distortion of the triangle at $T < T_N$ can happen
along arbitrary side of a triangle structure, we can expect three absorption
lines from the three domains shown in the insets. For field directions parallel
and perpendicular to one side of the triangular structure, two of three domains
will be in equivalent resonance conditions.

Red symbols in Figs.~\ref{fig1},~\ref{fig2} correspond to domain with the
distortion along the $[110]$ direction. Black and blue symbols throughout this
paper correspond to the resonances from two other domains with distortions
along the $[100]$ or $[010]$ directions. The $\omega(H_{res})$ dependence
measured for $\vect{H}\parallel[\bar{1}10]$ direction for $[110]$ domain (red
symbols on Fig.~\ref{fig2}) shows abrupt softening of one resonance frequency
at $H_c\approx 5.3$~T. This field corresponds to the spin reorientation
transition observed in Ref.~[\onlinecite{Yamaguchi_2010}]. The spin
reorientation can only be observed in the geometry with the static magnetic
field applied along the plane of one of the spin cycloids. If magnetic field is
applied perpendicular to the cycloidal plane (as in Fig.~\ref{fig1}) then two
other domains rotate continuously and no spin reorientation occurs.

For the $[110]$ domain the frequency-field dependence is
quasi-linear with field: $\nu = k\sqrt{H^2+\Delta^2}$. The
coefficient $k$ is noticeably smaller than the gyromagnetic ratio
$g\mu_B/h$ in two cases: i) at $\vect{H}\parallel[110]$ in full
studied field range (red circles in Fig. \ref{fig1}), and, ii) at
$\vect{H}\parallel[\bar{1}10]$ for fields higher than spin-flop
transition $H>H_c$ (red circles in Fig. \ref{fig2}, $H>H_c$). Such a
field dependence is typical for a planar spin structure with strong
``heavy axis'' anisotropy for vector $\vect{n}$ perpendicular to the
spin plane and for the field directed normal to ``heavy axis''
$(\vect{H} \perp \vect{c})$.\cite{Svistov_2009} The coefficient $k$ is defined by the
susceptibility anisotropy of the spin structure\cite{Andreev_1980,
Svistov_2009} : $k =
g\mu_B/h\sqrt{\chi_{\parallel}/\chi_{\perp}-1}$. In case of
CuCrO$_2$ the anisotropy of susceptibility is weak and, as a result,
the asymptotic derivative of $k$ is much smaller than the
paramagnetic value $g\mu_B/h$. For field direction
$\vect{H}\parallel[\bar{1}10]$ and for $H<H_c$ the branch is
expected to be quasi-linear with usual derivative limit of
$k=g\mu_B/h$ (red circles in Fig. \ref{fig2}, $H<H_c$)

For the $[100]$ and $[010]$ domains the spin plane rotates
monotonously with field. These modes are given by black-blue squares
in Figs.~\ref{fig1},\ref{fig2}. The accurate theoretical description
of frequency/field modes for such orientation of the spin plane
needs numerical calculations.  Nevertheless, we can expect that the
frequency/field mode for these domains will be situated between the
extreme cases described above: branches with $k=g\mu_B/h$ and $k =
g\mu_B/h\sqrt{\chi_{\parallel}/\chi_{\perp}-1}$.  The observed
frequency/field mode for these domains (black-blue squares in Fig.
\ref{fig1},\ref{fig2}) is in agreement with this statement, and the
``crossing'' with the $g=2$ line seems to be
natural\cite{Svistov_2009}.

%For domain $[110]$ the frequency-field dependence is quasi-linear
%with field: $\nu = k\sqrt{H^2+\Delta^2}$. The coefficient $k$ is
%noticeably smaller than the gyromagnetic ratio $g\mu_B/h$ at
%$\vect{H}\parallel[110]$ in full studied field range $(\Delta^2>0)$
%and at $\vect{H}\parallel[\bar{1}10]$ and fields higher than
%spin-flop transition $H>H_c$ $(\Delta^2<0)$. Such a field dependence
%is typical for a planar spin structure with strong ``heavy axis''
%anisotropy for vector $\vect{n}$ perpendicular to the spin plane and
%for the field directed normal to ``heavy axis'' $(\vect{H} \perp
%\vect{c})$.\cite{Svistov_2009}

\begin{figure}
\includegraphics[width=.9\columnwidth,angle=0,clip]{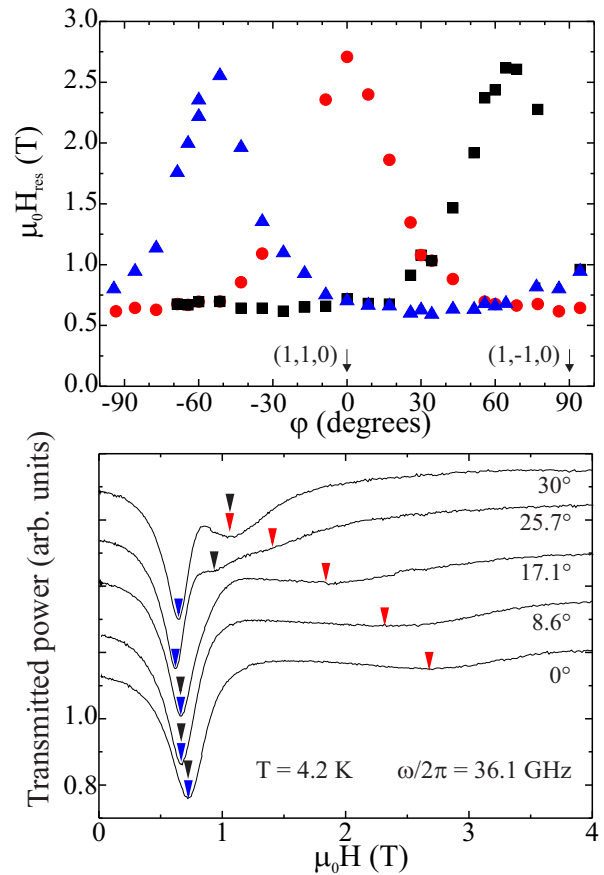}
\caption{(color online) Upper panel: Angular dependence of the
resonance fields for static magnetic field rotated within the plane
of the triangular structure (001) and at $\omega/2 \pi = 36.1$~GHz.
Red circles, black squares, and blue triangles correspond to the ESR
modes from three domains with distortions along [110], [100], and
[011], respectively. Lower panel: Examples of the ESR absorption
spectra for different directions of the external magnetic field.}
\label{fig3}
\end{figure}

Angular dependence of the resonance fields measured at $\omega/2\pi=36.1$ GHz
is shown in Fig.~\ref{fig3}. Magnetic field is rotated within the plane of the
triangular structure, i.e. within the(001)-plane. Red circles, black squares,
and blue triangles in the $\omega(H)$ spectra and red, black and blue arrows at
the absorption lines from the lower panel can be attributed to the resonances
from three magnetic domains. The angular dependence of resonance field
($H_{res}$) from every domain has 180 degrees periodicity. The angles at which
the resonance fields of different modes coincide follow a triangular geometry.
The spectrum in the lower panel of Fig.~\ref{fig3} taken at $\varphi =
0^{\circ}$ is equivalent to the spectrum in the lower panel of Fig.~\ref{fig1}
measured at the same frequency (36.1~GHz). In a similar way, the spectrum at
$\varphi = 30^{\circ}$ in Fig.~\ref{fig3} is similar to the spectrum at
36.7~GHz in Fig.~\ref{fig2}. Thus, the results in Fig.~\ref{fig3} establish a
continuous transformation of spectra in Fig.~\ref{fig1} to the spectra in
Fig.~\ref{fig2}. These dependencies allow to separate absorption features from
different domains.

\begin{figure}
\includegraphics[width=.9\columnwidth,angle=0,clip]{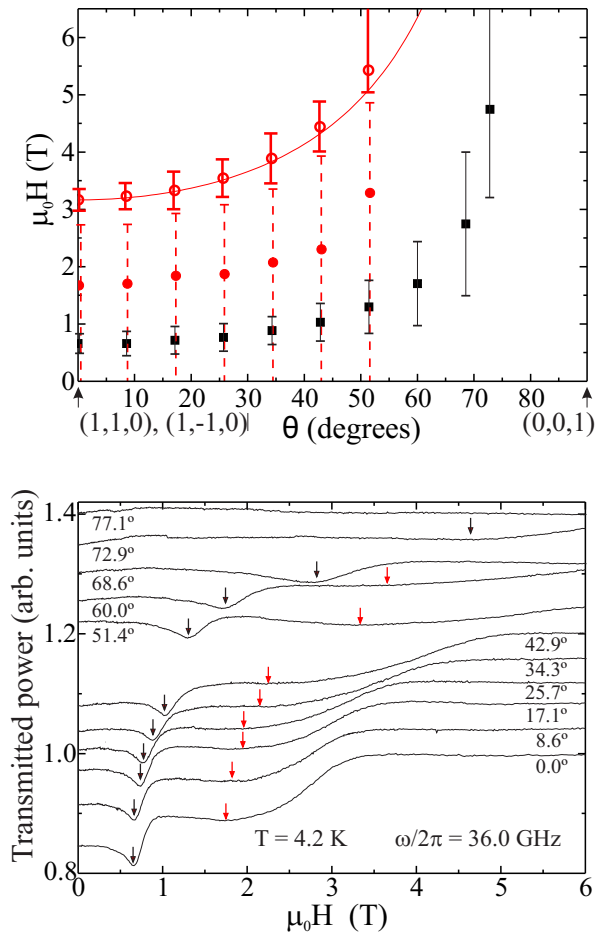}
\caption{(color online) Upper panel: (solid symbols) Angular
dependencies of the resonance fields for magnetic field rotated
within the $(\bar{1}10)$ plane and at $\omega/2 \pi = 36.0$~GHz;
(open symbols) angular dependencies of $H_{res}$ for the rotation of
the magnetic field within $(110)$ plane and at 89.7 GHz. The angles
are measured with respect to the trigonal plane. Red symbols
correspond to the resonances from the $[110]$ domains, black symbols
- from $[100]$ or $[010]$ domains. The error bars mark the line
width measured on the half level of the absorbed power.
 Lower panel: examples of the ESR absorption for
different directions of the applied magnetic field. $T=4.2$ K.}
\label{fig4}
\end{figure}

Fig.~\ref{fig4} shows the angular dependence of the resonance field for the
out-of-plane rotation of $\vect{H}$. Solid symbols in the upper panel
correspond to experiments with field rotated  within the $(\overline{1}10)$
plane. The resonance field initially increases with increasing angle and for
$\theta > 70^{\circ}$ rapidly exceeds our experimental field range. Since the
frequency in these experiments was near the gap of the low frequency branch
($\approx 33$~GHz) we can conclude, that for field directions perpendicular to
trigonal plane, the low frequency branch is almost independent on $H$. This
fact indicates strong anisotropy along the hexagonal axis in CuCrO$_2$.

Similar angular dependencies of the resonance positions were
obtained for the field rotated within $(110)$ plane. The
corresponding angular dependence of $H_{res}(\theta)$ at
$\omega/2\pi=89.7$ GHz for the $[110]$ domains is shown in
Fig.~\ref{fig4} with red open symbols. We can expect, that for such
field rotation and for $H<H_c$ the orientation of the $(110)$ spin
plane does not change. Therefore, in case of strong anisotropy along
the hexagonal axis, the resonance field will be defined by the field
projection on the hexagonal plane, i.e.:
$H_{res}=H_{res}(\theta=0)/\cos(\theta)$. This dependence is given
in Fig.~\ref{fig4} with a solid line and agrees well with the
experimental points.

\begin{figure}
\includegraphics[width=.9\columnwidth,angle=0,clip]{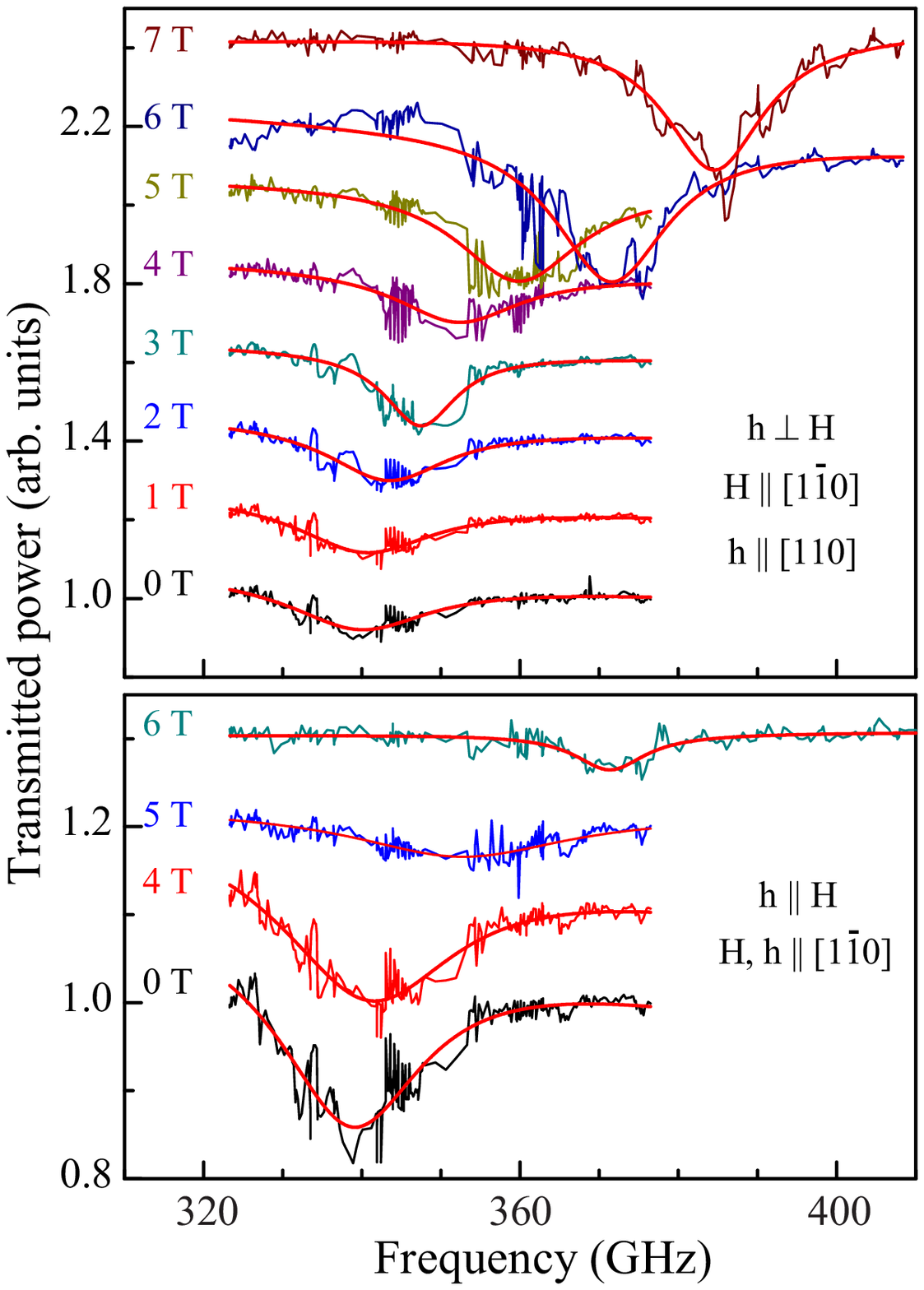}
\caption{(color online) Frequency dependence of the transmitted
power at different magnetic fields  measured in the quasioptical
geometry and for two polarizations of electromagnetic waves:
$\vect{h}\perp\vect{H}$ (upper panel) and $\vect{h}\parallel
\vect{H}$ (lower panel). The radiation propagates along the
crystallographic $[001]$ axis and perpendicular to the large plane
of the platelet sample. Thick solid lines shows the fits to the
spectra using the Lorenz line shape. The orientation of the magnetic
domains is shown in Fig.~\ref{fig5a}. $T=3$~K. } \label{fig5}
\end{figure}

\begin{figure}
\includegraphics[width=.9\columnwidth,angle=0,clip]{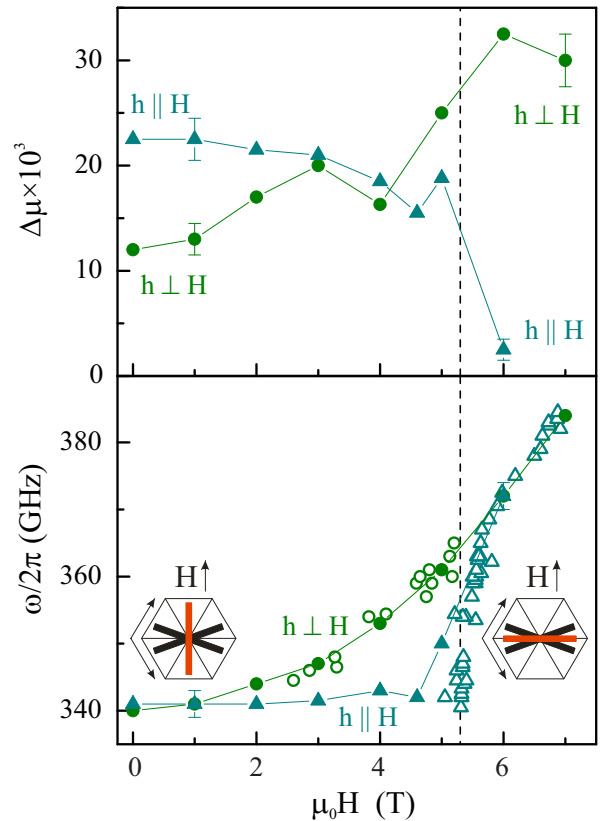}
\caption{(color online) Parameters of the high-frequency ESR mode in
CuCrO$_2$. The data were obtained for two different geometries as
indicated: $\vect{h}\perp\vect{H}$ (circles) and $\vect{h}\parallel
\vect{H}$ (triangles). In both cases the radiation propagates
perpendicular to the hexagonal plane and the ac magnetic field
$\vect{h}$ is within the plane. Upper panel: magnetic contribution,
lower panel: resonance positions. Solid symbols were obtained from
the frequency scans, open symbols - from the magnetic field scans at
a fixed frequency. Solid lines are to guide the eye, dashed line
indicates the position of the critical spin-flop field. The insets
in the lower panel show the suggested orientation of the magnetic
domains below and above $H_c$. $ T=3$~K.} \label{fig5a}
\end{figure}

The high frequency branch of the ESR spectra in CuCrO$_2$  was studied using
the quasi-optical technique. Transmitted power through the CuCrO$_2$
single-crystalline platelet at various magnetic fields $\vect{H}\parallel\
[1\bar{1}0]$ are shown in Fig.~\ref{fig5}. These frequency dependencies were
obtained by division of transmitted power measured at $T=3$~K by transmitted
power measured in the paramagnetic state, $T=30$~K ($T>T_N$). Such procedure
was used to separate the weak signal of the magnetic resonance absorption from
other contributions like standing waves within the sample. Fitting of the
absorption with Lorenz line shape is shown with a thick solid line. Magnetic
field dependencies of the parameters of the observed modes are shown in
Fig.~\ref{fig5a}. The measurements were performed in Voigt geometry with
$\vect{k} \perp \mu_0 \vect{H}$ and at two polarizations of electromagnetic
waves: $\vect{h}\perp\vect{H}$ and $\vect{h}\parallel \vect{H}$. Here
$\vect{k}||[001]$ is the wave vector of the electromagnetic radiation.

In the case $\vect{h}\perp\vect{H}$, as shown in the upper panel of
Fig.~\ref{fig5}, only the high-frequency mode of the $[100]$ and
$[010]$ domains is excited for $H<H_c$. After the spin flop
transition for $H>H_c$ the $[110]$ domains are rotated by 90 degrees
and they can be excited by the ac field $\vect{h}\perp\vect{H}$ as
well. This explains the increase in the mode intensity in high
magnetic fields as observed in the upper panel of Fig.~\ref{fig5}
and given as solid circles in the upper panel of Fig.~\ref{fig5a}.

In the geometry with $\vect{h}\parallel \vect{H}$ (lower panel of
Fig.~\ref{fig5}) both $[100]$ and $[010]$ domains are only weakly
excited because the excitation conditions are more favorable for
[110] domains. In this case the observed signal basically comes from
the $[110]$ domains which dominate the spectra. In low magnetic
fields the resonance frequency of these domains is practically
field-independent (solid triangles in the lower panel of
Fig.~\ref{fig5a}). After the spin flop transition for $H>H_c$ the
[110] domains cannot be excited in the geometry $\vect{H}||\vect{h}$
(see the right inset in Fig.~\ref{fig5a}). In this case only a weak
signal from [100] and [010] domains is observed. This is in
agreement with a suppression of the mode intensity as shown in the
upper panel of Fig.~\ref{fig5a} by solid triangles.

\section{Building and Control of Magnetic domains in C$\rm{u}$C$\rm{r}$O$_2$}

\begin{figure}
\includegraphics[width=.9\columnwidth,angle=0,clip]{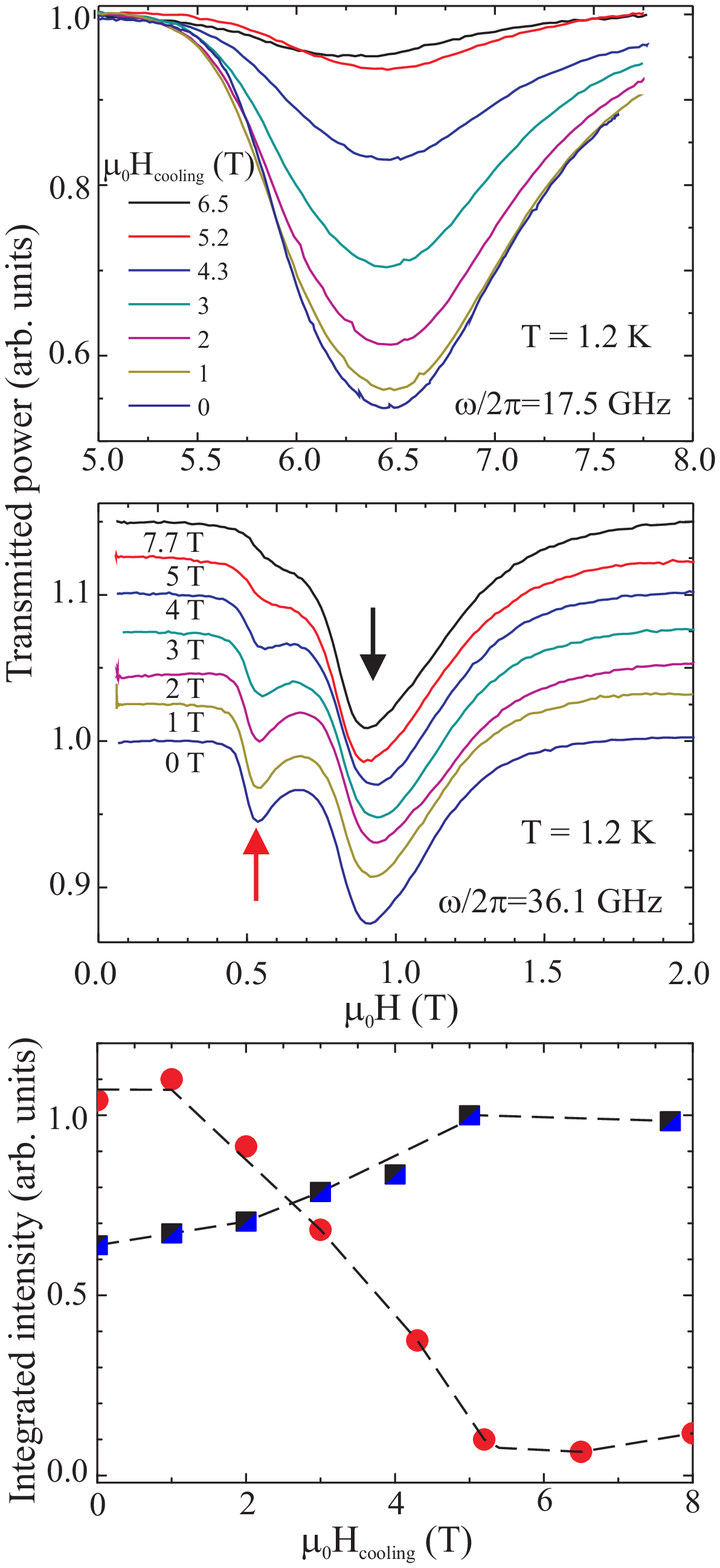}
\caption{(color online) ESR absorption lines measured for magnetic
field perpendicular to one side of the triangular structure
($\vect{H} \| [\bar{1}10]$) and at $T=1.2$~K. Top panel: absorption
line measured at 17.5~GHz which corresponds to the $[110]$ domains
and is seen for $H>H_c$. Middle panel: absorption lines of the
$[110]$ domains and $[100]$+$[010]$ domains measured at 36.1~GHz,
marked with red and black arrows respectively. The data were
obtained for a field-cooled sample at different fields
$H_{\textrm{cooling}}$. Bottom panel: the integral intensity of the
ESR lines as function of $H_{\textrm{cooling}}$ for $[110]$  domains
obtained from the curves in the top panel (red circles) and for
$[100]$+$[010]$ domains obtained from the middle panel (black
squares). Dotted lines are guides for the eye.} \label{fig7}
\end{figure}

\begin{figure}
\includegraphics[width=.9\columnwidth,angle=0,clip]{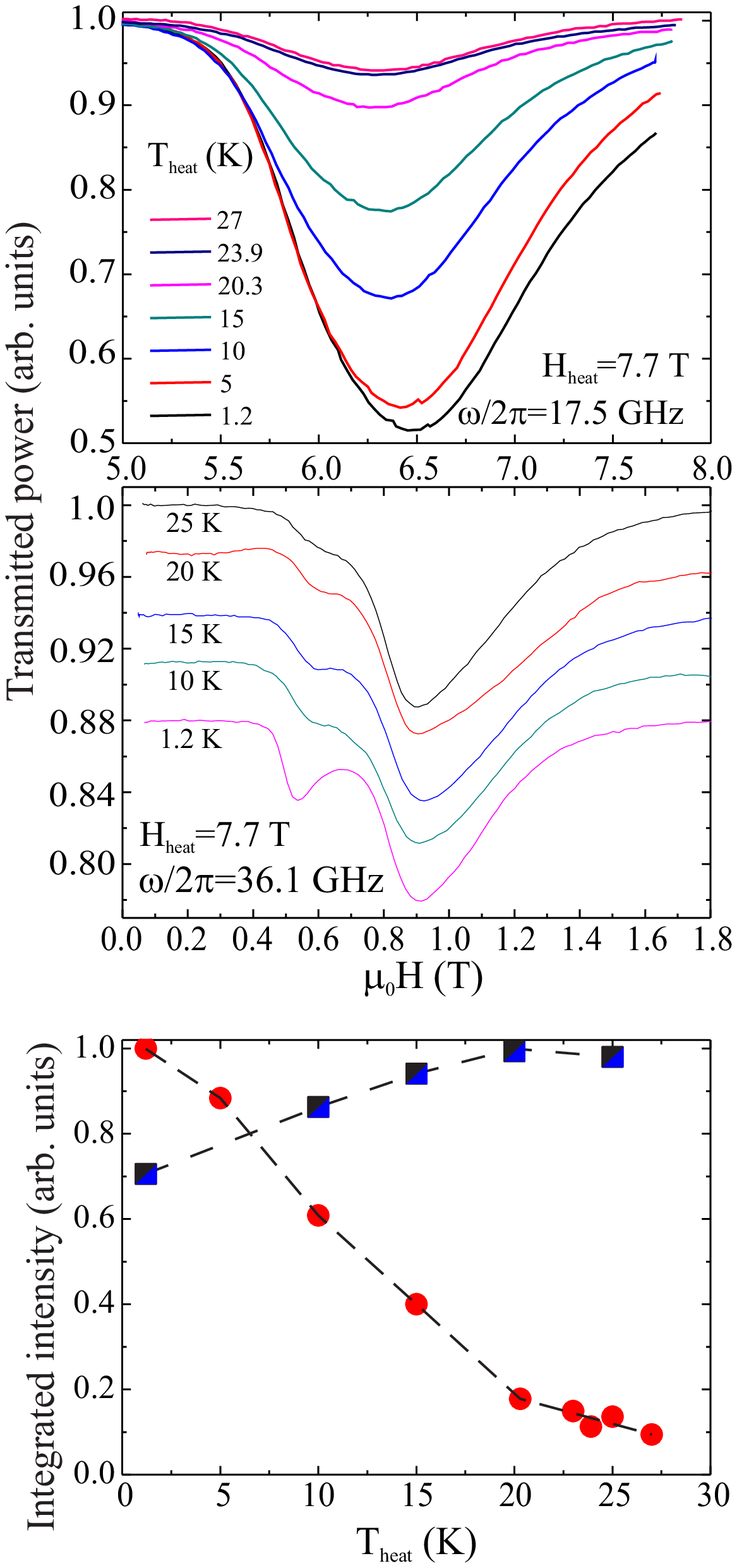}
\caption{(color online)  ESR absorption lines measured for magnetic
field perpendicular to one side of the triangular structure
($\vect{H}
\parallel [\bar{1}10]$) and at $T=1.2$~K. Top panel: absorption line measured
at 17.5 GHz which corresponds to the $[110]$ domains and is seen for
$H>H_c$. Middle panel: absorption lines of the $[110]$ domains and
$[100]$+$[010]$ domains measured at 36.1~GHz and marked with red and
black arrows respectively. The data were measured after annealing of
the sample at different $T_{\textrm{heat}}$ and at $\mu_0
H_{\textrm{heat}}=7.7$~T. Bottom panel: the integral intensity of
the ESR lines as a function of $T_{\textrm{heat}}$ for the $[110]$
domains obtained from the data of the top panel (red circles) and
for $[100]$+$[010]$ domains obtained from the middle panel (black
squares). Dotted lines are guides for the eye.} \label{fig8}
\end{figure}

As discussed above,  the ESR technique allows to recognize the
absorption modes originating from different domains. In this
sections we demonstrate that structural and magnetic domains are
strongly sensitive to thermal and magnetic history of the sample and
that magnetic domains are mobile already at temperatures close to
5~K.

To obtain the results presented in this chapter we repeated the
experiments within the geometry of Fig.~\ref{fig2}, i.e. at magnetic
fields perpendicular to one side of triangle structure:
$\vect{H}\parallel [\bar{1}10]$. The data have been obtained at the
lowest temperature of our spectrometer $T=1.2$~K and at two
frequencies of 36.1~GHz and 17.1~GHz. As discussed above, within
this geometry the spectra at lower frequency show a magnetic mode
above critical spin-flop field and they are sensitive to [110]
domains only. The spectra at 36.1 GHz show two modes at about 0.5~T
and 0.9~T (i.e below the spin-flop field) and they correspond to a
signal from [110] and ([100]+[010]) domains, respectively.
Therefore, analyzing the intensities of the corresponding modes as a
function of magnetic and thermal history, we obtain information
about relative population of different domains.

Figs. \ref{fig7},\ref{fig8} show the ESR spectra measured within the geometry
of Fig.~\ref{fig2} but for different magnetic and thermal history of the
sample. In the experiments presented in Fig.~\ref{fig7} the sample was cooled
in various static magnetic fields as indicated (field-cooled regime). In the
experiments presented in Fig.~\ref{fig8} the mobility of the domains was
investigated. The sample was initially cooled down to 1.2~K in zero magnetic
field, producing all three domains.  In a second step the sample was annealed
at various temperatures $T_{\rm{heat}}$ and at static magnetic field $\mu_0
H_{\rm{heat}}=7.7$~T applied along the $[\bar{1}10]$ direction. Such field
annealing lasted two minutes. In a following step first the temperature and
then the field were reduced to $T=1.2$~K and to $H=0$. Such treatment has been
done prior to every field scan shown in Fig.~\ref{fig8}.

First, we start the discussion with the results of the field-cooling
experiments, which are shown in Fig.~\ref{fig7}. The absorption lines measured
at 17.5~GHz and attributed to the domains with distorted $[110]$ side are shown
on the top panel of Fig.~\ref{fig7}. These absorption modes appear at magnetic
fields higher the spin-flop field: $H>H_c$. Clearly, the intensity of the
observed mode is strongly suppressed for the field-cooled sample.

The middle panel of Fig.~\ref{fig7} shows the spectra measured at
fields below the spin flop transition: $H<H_c$. The mode that is
marked by red arrow corresponds to the absorption line from $[110]$
domains and it obviously demonstrates the same behavior as the
spectra in the upper panel. In addition, the mode from two other
domains, $[100]$ and $[010]$, is observed in these experiments as
well. This mode is more intensive in this geometry and it is marked
by black arrow. The field cooling reduces the integral intensity of
absorption line from the $[110]$ domains and increases the intensity
from two other domains. This experiment shows, that a comparatively
small external field suppresses the energetically less favorable
domains and shifts the ESR intensity to two other domains. This
qualitative analysis is supported by the values of integrated
intensities of the ESR modes as given in the bottom panel of
Fig.~\ref{fig7}.

The absorption lines shown in Fig.~\ref{fig8} were obtained using
the cooling history as described above. Here the initial zero-field
cooling step produces all three domains. In this case, the spectra
taken at $T_{\mathrm{\rm{heat}}}=1.2$~K in Fig.~\ref{fig8} closely
follow the curves with $H_{\mathrm{cooling}}=0$ in Fig.~\ref{fig7}.
The subsequent annealing of the zero-field cooled sample in magnetic
field results in decrease of the intensity of absorption from
energetically not favorable $[110]$ domains and the transfer of the
ESR intensity to two other domains. Surprisingly, the domain
distribution changes even at $T_{\rm{heat}}\ll T_N$. Small change of
absorption line intensity was observable even after annealing at
$T_{\rm{heat}}=5$~K. The bottom panel of Fig.~\ref{fig8} show the
change in the integral intensities of the absorption lines from
different domains as a function of annealing temperature.

\section{Discussion}

\begin{figure}
\includegraphics[width=.9\columnwidth,angle=0,clip]{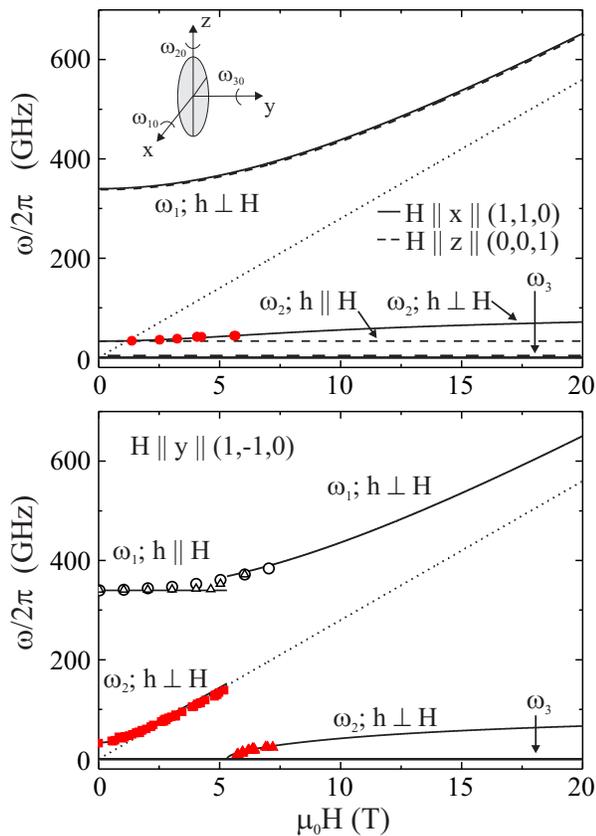}
\caption{(color online)  Theoretical frequency-field dependencies of
the ESR modes for field directions $\vect{H} \parallel [110]$ and
$\vect{H}
\parallel [\bar{1}10]$. The spectra are
computed following Ref.~[\onlinecite{Svistov_2009}] and using three
parameters $\omega_{10}/2\pi=340$~GHz, $\omega_{20}/2\pi=33$~GHz and
$H_{cy}=5.3$~T. The polarization conditions necessary for the
excitation of the ESR are given near the branches. Experimental
resonance frequencies $\omega(H_{\textrm{res}})$ for the domains
with the distortion of the triangular structure along the $[110]$
direction are shown with red solid symbols. Open black symbols show
the ESR $\omega_{\textrm{res}}(H)$ obtained from quasi-optical
experiments at $\vect{h}\perp\vect{H}$ (circles) and at
$\vect{h}\parallel \vect{H}$ (triangles). The dotted line
corresponds to a paramagnetic mode with $g=2$.} \label{fig6}
\end{figure}

The field and angular dependencies of the resonance frequencies as
observed at $T\ll T_N$ in CuCrO$_2$ can be consistently described by
the model of coplanar exchange spin-structure whose orientation in
space is defined by weak relativistic  interactions with external
field and by the crystal environment. The phenomenological
hydrodynamical theory of macroscopic dynamics of magnets with
dominant exchange interactions was developed in
Ref.~[\onlinecite{Andreev_1980}]. The application of this theory to
the coplanar magnetic structures was described in
Ref.~[\onlinecite{Svistov_2009}] and will be used in the following
discussion. The anisotropic part  of the energy of magnetic
structure of CuCrO$_2$ can be written as:
\begin{eqnarray}
U=-\frac{\chi_{\parallel}-\chi_{\perp}}{2}(\vect{n}\vect{H})^2+\frac{1}{2}(An_{x}^2+Bn_{y}^2),
\label{eqn:energy}
\end{eqnarray}
\noindent where $\vect{n}$ is a unit vector perpendicular to the
spin plane; $\chi_{\parallel}$ and $\chi_{\perp}$ are the
susceptibilities of the planar structure parallel and perpendicular
to $\vect{n}$ and they are defined by exchange interactions; $A$ and
$B$ are the anisotropy constants. For the case of CuCrO$_2$:
$\vect{z}\parallel[001]$, $\vect{y}\parallel[\bar{1}10]$,
$\vect{x}\parallel[110]$, $A < B < 0$ and $\chi_{\parallel} >
\chi_{\perp}$. The minimum of energy at $\vect{H}\parallel\vect{x}$
is realized for the structure with $\vect{n}
\parallel \vect{x}$. At field directions $\vect{H}\parallel\vect{z}$
and $\vect{H}\parallel\vect{y}$ a spin reorientation take place at
critical fields $H_{cz}^2=-A/(\chi_{\parallel}-\chi_{\perp})$ and
$H_{cy}^2=(B-A)/(\chi_{\parallel}-\chi_{\perp})$, respectively. For
fields below $H_c$  the spin plane is oriented by crystal
anisotropy: $\vect{n}\parallel\vect{x}$. At $H>H_c$ the spin plane
is oriented by field: $\vect{n}\parallel\vect{H}$. The resonance
frequencies of acoustic modes of planar spin structures can be
obtained in the frame of Lagrangian formalism~\cite{Andreev_1980}
using the potential energy in form of Eq.~(\ref{eqn:energy}). The
resonance frequencies at zero magnetic field are:
$w_{10}^2=\gamma(-A)/\chi_{\perp}$;
$w_{20}^2=\gamma(B-A)/\chi_{\perp}$, and $w_{30} =0$, where $\gamma$
is the gyromagnetic ratio. These frequencies correspond to
oscillations of the spin structure around three axes as
schematically shown in the inset to Fig.~\ref{fig6}. The zero energy
oscillation ($w_{3}$) indicates the degeneracy of the ground state
with respect to rotations of the structure around the vector
$\vect{n}$. The experimental values of $w_{10}/2\pi=340$ GHz and
$w_{20}/2\pi=33$ GHz and the spin-flop field $H_{cy}=5.3$ T at
$\vect{H}\parallel\vect{y}$ obtained here are in a good agreement
with the results given in Ref.~[\onlinecite{Yamaguchi_2010}]. The
theoretical curves describe well the experimental points. A small
deviation $(~7\%)$ of experimental points ascribed to domain
$[110]$ at $\vect{H}\parallel[\bar{1}10]$, $H<H_c$, from theoretical
dependence (See Figs.~\ref{fig2},~\ref{fig6}), is most probably due
to higher order anisotropy terms  in Eq.~(3), which are not
considered in the discussed model.

The values of $ w_{20}$ and $H_{cy}$ allow to define the
susceptibility anisotropy of the spin structure:
$\eta=(\chi_{\parallel}-\chi_{\perp})/\chi_{\perp}=(\omega_{20}/\gamma
H_{cy})^2=0.045~\pm 0.03$. This small anisotropy is a result of
close similarity of the magnetic structure of CuCrO$_2$ to the
regular 2D triangular structure, for which, if we neglect
fluctuations, $\chi_{\parallel}$ is equal to $\chi_{\perp}$.

From the present results we can evaluate the spin-flop field for
$\vect{H}\parallel\vect{z}$ as: $H_{cz}=H_{cy}(w_{10}/w_{20})\approx
55$ T. This value exceeds by far our available experimental range,
but is still much below the expected saturation field of CuCrO$_2$.
The latter can be estimated using the susceptibility value
$\chi=0.006$ emu/mol,\cite{Kimura_2008} as: $H_{sat}\approx
M_{sat}/\chi \approx 280$ T.

Theoretical field dependence of the resonance frequencies for field
directions $\vect{H}
\parallel [110]$ and $\vect{H} \parallel [\bar{1}10]$ are shown in
Fig.~\ref{fig6}. The spectra are defined by three parameters
$\omega_{10}$, $\omega_{20}$ and $H_{cy}$ as given above. The
experimental resonance frequencies for the domains with distortion
of the triangular structure along $[110]$ direction are shown in the
same figure with red symbols. Open black symbols in Fig.~\ref{fig6}
show the high frequency excitation branches from all three domains.

The excitation conditions of the magnetic modes are indicated at the
theoretical curves in Fig.~\ref{fig6}. The high frequency magnetic
field $\tilde{h}$ is directed along the vector $\omega$ denoting the
corresponding oscillation of the spin plane~\cite{Andreev_1980}. The
high frequency branch at $\vect{H}
\parallel [\bar{1}10]$ and $H<H_{cy}$ can be excited by
$\vect{h}\parallel\vect{H}$ and for $H>H_{cy}$ by $\vect{h}\perp
\vect{H}$. Such polarization excitation conditions agrees well with
the present experiment (see Fig.~\ref{fig5}).

The experimental study of the magnetic domain distribution in the single
crystals of CuCrO$_2$ shows that their relative size depends on the magnetic
history of the sample. The sample cooled at zero magnetic field has three
domains of comparable volume. Both, the field cooling and the field annealing
reduces the volume of the domains which are oriented unfavorably with respect
to static field. The data demonstrates that the rebuilding of the domains takes
place at temperatures much below the N\'{e}el temperature. Such behavior
indicates an anomalously large mobility of magnetic domain walls in CuCrO$_2$.
The observed effect needs further investigations in order to test the magnetic
structure during the rebuilding of the domains. In particular, it is necessary
to prove experimentally that the wave vector direction within the domains
agrees with the in-plane anisotropy.

We note that for each magnetic domain with a specific orientation of the spin
plane two different directions of the chirality vector exist. According to
Refs.~[\onlinecite{Kimura_2009,KimuraJ_2009}] and due to magnetoelectric
coupling in CuCrO$_2$ these magnetic domains can be switched by external
electric fields. Such high sensitivity of the magnetic structure in CuCrO$_2$
to external effects makes this system very attractive from the experimental
point of view.

\section{Conclusions}

We performed detailed electron-spin-resonance studies of a
frustrated triangular quasi-two-dimensional antiferromagnet
CuCrO$_2$. The results of low temperature ESR experiments are
well described in the frame of phenomenological model of coplanar
spin structure with biaxial anisotropy. Our calculations reproduce
well the experimentally observed changes of the character of the
eigenmodes at the spin-flop transition. The observed splitting of
the modes is attributed to resonances from different domains which
can be controlled by annealing of the sample in magnetic field.

CuCrO$_2$ is an example of a frustrated triangular
quasi-two-dimensional antiferromagnet with spin $S=3/2$. This
material thus occupies an intermediate position between the systems
with large spin which are intensively studied experimentally and
theoretically and the systems with $S=1/2$, for which the
experimental objects are still far from being a model for theory.

\acknowledgements

We thank V. I. Marchenko for useful discussion. This work is supported by
Russian Foundation for Basic Research, Program of Russian Scientific Schools
(Grants 12-02-00557-a, 10-02-01105-a, 11-02-92707-IND-a, 12-02-31220 mol\_a)
and by Austrian Science Funds (I815-N16, W1243).


\begin{thebibliography}{9}

\bibitem{Kawamura_1985} H.~Kawamura and S.~Miyashita, J. Phys. Soc. Jpn. {\bf 54}, 4530 (1985).
\bibitem{Korshunov_1986} S.~E.~Korshunov. J. Phys. C: Solid State Phys. {\bf 19}, 5927 (1986).
\bibitem{Anderson_1987} P.~W.~Anderson, Science {\bf 235}, 1196  (1987).
\bibitem{Plumer_1990} M.~L.~Plumer, A. Caille, Phys. Rev. B {\bf 42}, 10388 (1990).
\bibitem{Chubukov_1991} A.~V.~Chubukov and  D.~I.~Golosov, J. Phys.: Condens. Matter {\bf 3}, 69 (1991).
\bibitem{White_2007} Steven~R.~W.~White and A.~L.~Chernyshev, Phys. Rev. Lett {\bf 99}, 127004 (2007).
\bibitem{Kadowaki_1990} H. Kadowaki, H. Kikuchi and Y. Ajiro J. Phys.:  Condens. Matter {\bf{2}}, 4485-4493 (1990).
\bibitem{Poeinar_2009} M. Poienar, F. Damay, C. Martin, V. Hardy, A. Maignan, and G. Andre, Phys. Rev. B {\bf{79}}, 014412 (2009).
\bibitem{Kimura_2009} K. Kimura, H. Nakamura, S. Kimura, M. Hagiwara, and T. Kimura, Phys. Rev. Lett. {\bf{103}}, 107201 (2009).
%\bibitem{Marchenko 2013} V. I. Marchenko  (2013).
\bibitem{Yamaguchi_2010} H. Yamaguchi, S. Ohtomo, S. Kimura, M. Hagiwara, K. Kimura, T. Kimura, T. Okuda, and K. Kindo, Phys. Rev. B {\bf{81}}, 033104 (2010).
\bibitem{Beznos_2009} B. V. Beznosikov, K. S. Alexandrov, J. Str. Chem. {\bf{50}}, 108-113 (2009).
\bibitem{KimuraJ_2009} K. Kimura, T. Otani, H. Nakamura, Y. Wakabayashi, and T. Kimura, J. Phys. Soc. Jpn.  {\bf{78}}, 113710 (2009).
\bibitem{Frontzek_2012} M. Frontzek, G. Ehlers, A. Podlesnyak, H. Cao, M. Matsuda, O. Zaharko, N. Aliouane, S. Barilo, S. V. Shiryaev, J. Phys.:  Condens. Matter {\bf{24}}, 016004 (2012).
\bibitem{Poeinar_2010} M. Poienar, F. Damay, C. Martin, J. Robert, and S. Petit, Phys. Rev. B {\bf{81}}, 104411 (2010).
\bibitem{Soda_2009} M. Soda, K. Kimura, T. Kimura, M. Matsuura, K. Hirota, J. Phys. Soc. Jpn.  {\bf{78}}, 124703 (2009).
\bibitem{Seki_2008} S. Seki, Y. Onose, and Y. Tokura, Phys. Rev. Lett. {\bf{101}}, 067204 (2008).
\bibitem{volkov_infrared_1985} A.~A. Volkov, Yu.~G. Goncharov, G.~V. Kozlov, S.~P. Lebedev, and A.~M.
Prokhorov., Infrared Phys. \textbf{25},369 (1985).
\bibitem{pimenov_prb_2005} A.~Pimenov, S.~Tachos, T.~Rudolf, A.~Loidl, D.~Schrupp, M.~Sing,
R.~Claessen, and V.~A.~M. Brabers. Phys. Rev. B \textbf{72}, 035131
(2005).
\bibitem{ivannikov_prb_2002} D.~Ivannikov, M.~Biberacher, H.-A. Krug~von Nidda, A.~Pimenov,
A.~Loidl, A.~A. Mukhin, and A.~M. Balbashov. Phys. Rev. B \textbf{65}, 214422 (2002).
\bibitem{Andreev_1980} A. F. Andreev and V. I. Marchenko Usp. Fiz. Nauk {\bf{130}}, 39, (1980).
\bibitem{Svistov_2009} L. E. Svistov, L. A. Prozorova, A. M. Farutin, A. A. Gippius, K. S. Okhotnikov,
A. A. Bush, K. E. Kamentsev, and \'E. A. Tishchenko, JETP {\bf 135}, 1151 (2009) or see attachment in A.A. Bush, N. B\"{u}ttgen, A.A. Gippius, V.N. Glazkov, W. Kraetschmer, L.A. Prozorova, L.E. Svistov, A.M. Vasiliev, A. Zheludev and A.M. Farutin, e-print arXiv:1304.4728.
\bibitem{Kimura_2008} K. Kimura, H. Nakamura, K. Ohgushi, and T. Kimura, Phys. Rev. B. {\bf{78}}, 140401 (2008).
%\bibitem{Rastelli_1986} E. Rastelli and A. Tassi, J. Phys. C: Solid State Phys.  {\bf{19}}, L423 (1986).
%\bibitem{Rastelli_1987} E. Rastelli and A. Tassi, J. Phys. C: Solid State Phys.  {\bf{20}}, L303 (1987).
%\bibitem{Rastelli_1988} E. Rastelli and A. Tassi, J. Phys. C: Solid State Phys.  {\bf{21}}, 1003 (1988).

\end{thebibliography}
\end{document}